# Ultrafast Laser Nanostructured ITO Acts as Liquid Crystal Alignment Layer and Higher Transparency Electrode

A. Solodar,[1] A. Cerkauskaite [2], R. Drevinskas [2], P. G. Kazansky [2] and I. Abdulhalim [1,3]

[1]Ben Gurion University of the Negev, Department of Electro-Optics Engineering and The Ilse Katz Institute for Nanoscale Science and Technology, Beer Sheva 84105, Israel.
[2]Optoelectronics Research Centre, University of Southampton, Southampton, SO17 1BJ, United Kingdom.
[3]SHARE - NEW CREATE Programme, School of Materials Science and Engineering, Nanyang Technological University, Singapore, 639798.

E-mail: asisolodar@gmail.com

## Abstract

Electrodes with higher transparency that can also align liquid crystals (LCs) are of high importance for improved costs and energy consumption of LC displays. Here we demonstrate for the first time alignment of liquid crystals on femtosecond laser nanostructured indium tin oxide (ITO) coated glass exhibiting also higher transparency due to the less interface reflections. The nano paterns were created by fs laser directlly on ITO films without any additional spin coating materials or lithography procces. Nine regions of laser-induced nanostructures were fabricated with different alignment orientations and various pulse energy levels on top of the ITO. The device interfacial anchoring energy was found to be $1.063 \times 10^{-6}\ J/m^2$, comparable to the anchoring energy of nematic LC on photosensitive polymers. The device exhibits contrast of 30:1 and relaxation time of 330ms expected for thick LC devices. The measured transparency of the LC device with two ITO nanograting substrates is 10% higher than the uniform ITO film based LC devices. The alignment methodology presented here paves the way for improved LC displays and new structured LC photonic devices.

## 1. Introduction

Liquid crystal (LC) devices are widely used as building blocks of many electro-optical systems including linear polarization rotators, dynamical wave plate retarders, and pixilated devices for displays, spatial light modulators, and tunable filters [1-3]. A typical LC device consists of a thin LC layer situated between a pair of indium tin oxide (ITO) films, deposited on top of glass substrates, which serve as transparent conducting electrodes. In the absence of an external electric field, the director of the liquid-crystal molecules strives to orient in a specific direction according to the alignment layer at the electrodes surface. This effect plays a key role in liquid crystal devices, as it is determining the speed, the pretilt angle, the viewing angle as well as an operational mode, which typically can be parallel, anti- parallel or twisted nematic (TN) as depicted in figure 1. Therefore, precise alignment of the LC molecules is required for high quality components.

Several methods have been used to form the surface alignment layer. The most common alignment technique in liquid crystal industry comprises unidirectional mechanical rubbing of thin polyimide coating. These thin films are spin coated and then cured at appropriate temperature according to the polyimide type. Thereafter, the cured films are rubbed with a velvet cloth; producing micro or nano grooves along the rubbing direction to which the LC molecules are then aligned. However, the rubbing method has some shortcomings, due to electrostatic charge accumulation, surface damage and particle dust generation, as a result of the mechanical contact interaction between the cloth and the surface during the rubbing process [4].

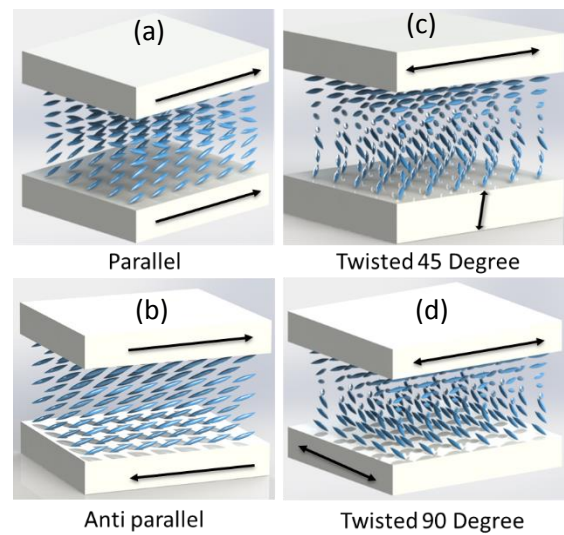

**Figure 1.** Some of the common LC alignment modes. (a) Parallel, (b) anti-parallel, (c) 45 degree twisted nematic (TN) and (d) 90 degree TN LC cell.

Liquid crystal molecules can also be aligned by noncontact methods[5], such as photo alignment by illumination with ultraviolet polarized light of photosensitive polymers, chemically by surface active agents (surfactants) or by ion beam irradiation. However, these techniques are not widely implemented in the industry, because of insufficient time stability, lower anchoring energy and requirement of expensive vacuum equipment [6,7].
Photo-alignment using polarized blue light was also demonstrated in our group on solid thin films of chalcogenide glasses [8-11].

In recent years, new alignment methods were investigated, utilizing nanomaterials structures based on Berreman's groove theory [12]. Taichi K. et al. [13], demonstrated homogeneous, tilted, and homeotropic alignment of nematic liquid crystal molecules using nanometer-sized ultrafine grooves fabricated on ITO substrates using electron beam lithography. Nanoscale patterns can also be employed for LC alignment by nanoimprint lithography as suggested by Yan Jun Liu et al.[14]. According to this method, ultrafine 50 nm line and space nanograting patterns formed by the imprinting process was used in order to assemble TN LC cell. Additional nanostructured alignment method was proposed by Woo-Bin Jung et al.[15], where they used tilted ITO line patterns in order to reduce switching time effects. A solution-derived alignment method was suggested by Hong-Gyu Park et al.[16], in which they demonstrated a homogeneous self-alignment of LCs on nanocrystalline $SnO_2$ solution, spin coated on ITO-glass substrates and annealed at temperature above 400 C for 1 hour. Another technique of interests is the LC alignment on the nanocolumnar structure produced by the oblique angle deposition technique [17]. Despite the great potential of the methods mentioned above, their fabrication process is still quite complicated and requires several steps in the fabrication process such as photoresist spin coating, lithography process, mould structures design, heating, etc.

In this work, we demonstrate a direct writing of nano-structured patterns on ITO coated glass surface by ultrafast femtosecond laser, as an alternative method for LC alignment layers. The patterned ITO electrodes are shown to be with higher transparency due the minimization of interface reflections; hence this methodology might find applications also in lowering the energy costs of optoelectronic devices. One of the most fascinating aspects of this technique is the ability to induce tunable structures with subwavelength periodicities at different orientations and so opens a new methodology to produce structured liquid crystal devices[18-19].

## 2. Experimental

### 2.1 Device fabrication process

During this research we built two LC devices, each comprises of 17x15x2.5 mm$^3$ ITO coated glass substrates (Think Film Devices Inc. California, USA) on either side of the device, with 145±10 nm thickness

The substrates were initially cleaned by ultrasonic bath with three steps process: deionized (DI) water, isopropanol (IPA) and at last with acetone, where each stage lasts about 20 minutes at a temperature of 50 degrees. Then cotton wipers with cleaning solvents were used in order to remove any unwanted materials on the surface of the glasses. In this process the surface is gently rubbed with cotton wipers dipped in acetone and IPA alternatively. This cleaning process was repeated until the substrates were clean. The first device consists of nine squared nanostructured regions on one glass substrate, as depicted in figure 2. The square regions separated by a uniform ITO regions, where the dimension of each square is 2mm x 2mm. The nanostructured regions were created by femtosecond laser system PHAROS (Light Conversion Ltd.) operating at a wavelength of 1030 nm with the repetition rate of 200 kHz and the pulse duration of 300fs. The squares were printed by scanning the sample at the speed of 1 mm/s with the interline distance of 2 µm.

It is important to emphasize that the nanogrid patterns during this work generated strictly on the ITO films by fs laser without any additional lithographic process, spin coating materials or rubbing.

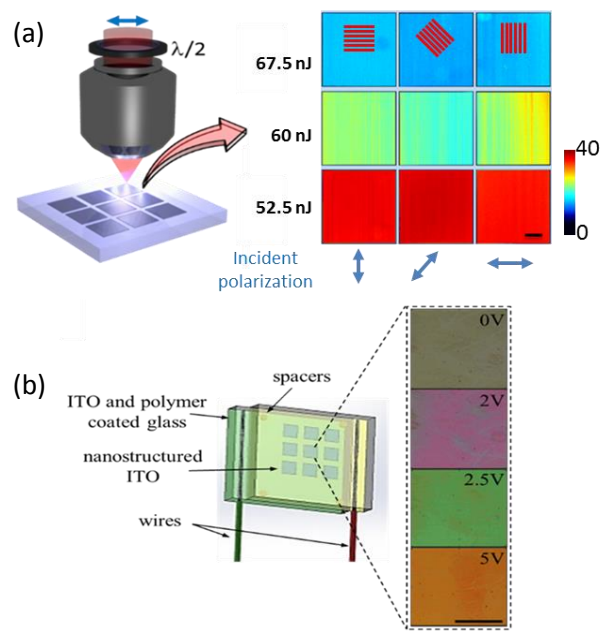

**Figure 2.** (a) Femtosecond laser nano structuring of ITO coated glass towards the fabrication of anisotropic surface elements. Nine regions were imprinted under different writing conditions. The nanostructure orientation (red lines) were imprinted by linearly polarized incident laser beam (blue arrows), controlled by rotating the half-wave plate (λ/2).The colours represents different retardation levels, which was imaged by quantitative birefringence measurement system. (b) Schematic of the constructed device. Inset shows polarized microscope images of a central region at different voltages for $45^0$ twisted mode. Scale bars are 0.5 mm.

Each row of the square nanostructure matrix was fabricated under different pulse energy of: 67.5nJ, 60nJ and 52.5nJ (from top to bottom) focused by 0.16 numerical aperture (NA) objective. Different pulse energies influence on the depth of the nanogrid patterns, causing various retardation levels at the specific regions (figure 2a). The orientations of the nanograting at each column was changed from horizontal to diagonal and to vertical (figure 2a), in order to achieve different alignment directions which, in turn, will compound three different surface modes as described in figure 1.

The generated ITO nanograting patterns on the glass surface are illustrated in the images taken by scanning electron microscope (SEM) and atomic force microscope (AFM) in figure 3 and figure 4, respectively.

The second ITO glass substrate, besides the aforesaid cleaning process, was additionally cleaned by ultra violet ozone cleaning system (UVOCS) for 50 minutes in order to achieve high hydrophilic surfaces with a small contact angle during the spin coating of the polymer alignment layer. Then the ITO substrate was spin coated with a sensitive polymer layer ROP - 108 (approximately 30nm thick), obtained from Rolic-Switzerland. After coating, the substrate was baked 15 minutes at $180^0$C and then slowly cooled down to room temperature. Thereafter, we used linearly polarized ultra-violet (UV) light of power density of 0.5mW/cm$^2$ for 20 minutes and then sandwiched together the two substrates with 10 micron spacers. During the construction a special emphasis was considered on the thickness variation across the cell in order to reduce to minimum any wedge formations that can cause non-uniform retardation and response time.

After assembly the gap was filled by capillary suction with E-7 LC (Merck). Finally, two electrodes were connected to the substrates as illustrated by figure 2b. By using these surface treatments, nine different squared alignment regions, which are equivalents to those shown in figure 1, were achieved. That is, each region in the matrix of squares (figure 2a) represents different LC alignment cell which is a proof that the LC molecular director follows the grating lines directions which act as the easy axis. Hence a homogeneously aligned device is obtained (parallel or anti-parallel geometry), a 45 degrees twisted device and a 90 degrees twisted device.

The second device was constructed completely from two ITO nanograting pattern glass substrates. In order to achieve faster laser material processing, the ITO substrates were irradiated by femtosecond laser system operating at a wavelength of 1030 nm with the repetition rate of 500 kHz and the pulses with duration of 400fs focused via 0.03 NA focusing lens. The scanning speed and the interline distance were fixed at 1 mm/s and 4 μm. The nanograting patterns on two ITO substrates fabricated under pulse energy of 0.76μJ and 0.68μJ respectively, over 12x12mm area (Figure 3a). Figure 3b describes SEM images at the border between the uniform ITO and nanograting region, while figure 3c and figure 3d illustrate the border region of the second device constructed with 2 nanogrid substrates and rotated by $45^o$ under polarized microscope. Without voltage bias (figure 3c) the aligned liquid crystal molecules at the nanogrid region exhibit maximum transmission intensity, while at the uniform ITO region the light was blocked by the polarizers. Upon voltage activation (figure. 3d) the aligned LC molecules at the nanogrid region has been stimulated by the applied electrical field. As a result the reflected spectrum colour at the nanogrid region has been changed, depending on the voltage, while the ITO uniform region remains color less.

The two substrates sandwiched together with parallel / anti-parallel orientation alignment, separated by 10μm spacers. Then LC BL036 was inserted into the cell in the isotropic phase by capillary action.

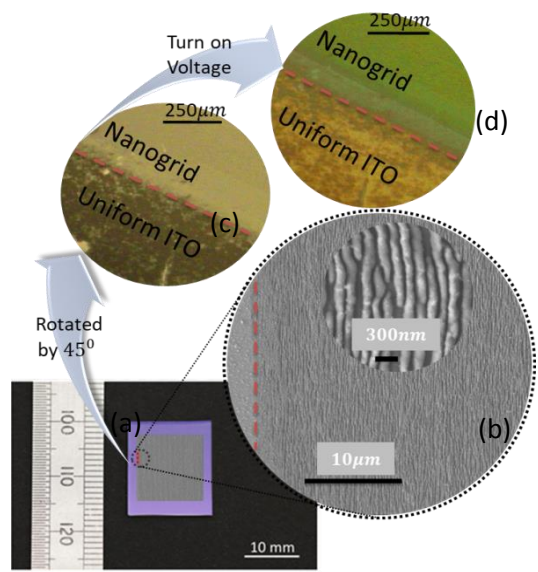

**Figure 3.** ITO grid substrate (a) and SEM images of nanostructured region (b). Red dashed line represents the border between uniform ITO surface (left) and nanograting patterns (right). Alignment behavior of the second device (rotated by 45$^o$) with two nanogrid substrates under crossed polarizers without voltage (c) and under bias of 8V, 100Hz and 1V offset (d).

### 2.2 Characterization

The Extinction and alignment states of the devices were examined under polarizing microscope (Olympus mx50) between crossed polarizers in transmission mode. The linearly polarized light (P) was focused by x10 objective onto the 9 regions of interest at the first device as illustrated in figure 5a. Under crossed polarizers, three different configurations are possible as illustrated in figure 5b. First, the device was aligned with its polymer coated alignment axis ($P_{ly}$) as well as a grid alignment axis ($P_{gr}$), along the polarizer (P) axis for the case of parallel / anti-parallel alignment as depicted in figure 5b. In the case of twisted $45^0$ alignment, the $P_{gr}$ axis was at 45 degrees respect to the analyzer axis, while the orientation of the $P_{ly}$ axis was parallel to the P axis. In the last case of twisted $90^0$ alignment, the $P_{gr}$ and $P_{ly}$ are parallel to A and P axis, respectively.

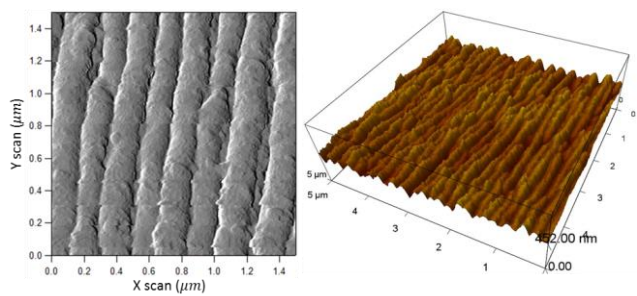

**Figure 4.** Nanogrid alignment structures. AFM tunnel patterns generated by fs laser on ITO film. Nanogrid pattern was written with fs laser under pulse energy of 0.76μJ.

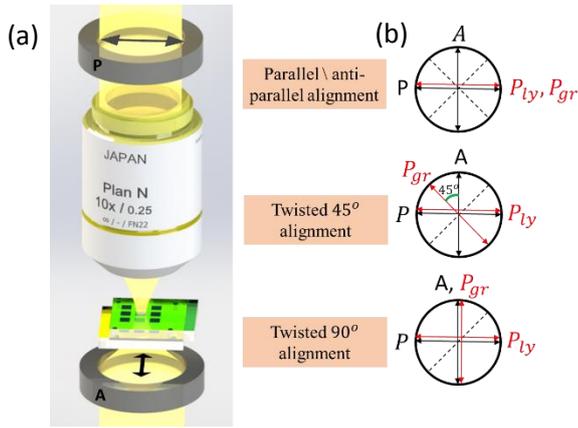

**Fig. 5** Illustration of LC device under the polarization microscope (a) between the crossed polarizer (P) and the analyzer (A). The polymer alignment direction, transmission axis of the polarizer, transmission axis of the analyzer and the direction of the nanogrid patterns are labeled as $P_{ly}$, P, A, and $P_{gr}$, respectively (b).

The transmission spectrum resulting from nine different nanogrid alignment regions are analyzed to deduce the performance of the device. The optical setup to produce transmission spectrum properties is illustrated in figure 6. The device was placed between crossed polarizers, where the polymer alignment axes is rotated at $45^0$ with respect to the polarizer. A collimated fiber white light source is passed through a linear polarizer and propagated through the regions of interest in the device. The transmitted light from the device is collected by x5 objective and focused onto the spectrometer fiber. The LC device is controlled by a function generator, which is automated with a spectrometer through LabView software.

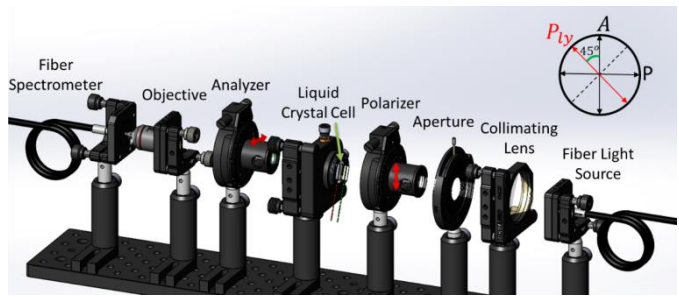

**Figure 6.** Schematic representation of the electro optical setup to determine transmission spectrum of the LC on the nanograting regions. The setup comprises a fiber light source collimated using a lens with focal length f=8cm, polarizer, aperture, the LC device attached to the 6-axis kinematic mount, an analyzer, a x5 focusing microscope objective, an output fiber connected to spectrometer (StellarNet Inc.).

## 3. Results and discussion

For the first proof of concept, we used polarizing microscope in order to examine the different alignment regions as illustrated in figure 5. It is noteworthy, that the molecules align themselves with the direction of the nanogrid tunnels from the one side (figure. 4a), and along the polymer alignment axis from the other side of the device. In the case of parallel\anti-parallel alignment, under crossed polarizers, the light transmitted through the device did not change its polarization direction, so that the light which reaches the analyzer cannot pass through it. In the case of twisted $45^o$ alignment device the incoming linear polarized light coincident with the polymer substrate alignment direction. Therefore, the linear polarized light entering through the polymer coated substrate will follow the twist and be rotated $45^o$ by the time it reaches the nanogrid patterned substrate. Upon transmission, the polarization direction will be rotated by $45^o$ so that the outcoming light will partially pass through the analyzer. According to the same principle, an incoming linearly polarized light transmitted through the twisted $90^o$ alignment region will follow the twist of the LC director. This leads to a rotation of the output polarization state along the analyzer (rotation of $90^o$), which ensures a total transmission of the outcoming light through the analyzer. Figure 7 shows a variety of output extinction states, obtained by the microscope camera under crossed polarizers.

Then, the reflection intensity was captured from the central region by the microscope camera for various applied voltages as illustrated in figure 2b, while the device with the polymer alignment axis oriented at 45 degrees relative to polarizer axis. Characterization of the LC device was achieved by spectral transmission measurements of the nine regions between crossed polarizers and an angle of $45^0$ between the polymer alignment axis and the polarizer as depicted in figure 6. The measured transmission spectra at various applied voltages are shown in Figure 8.

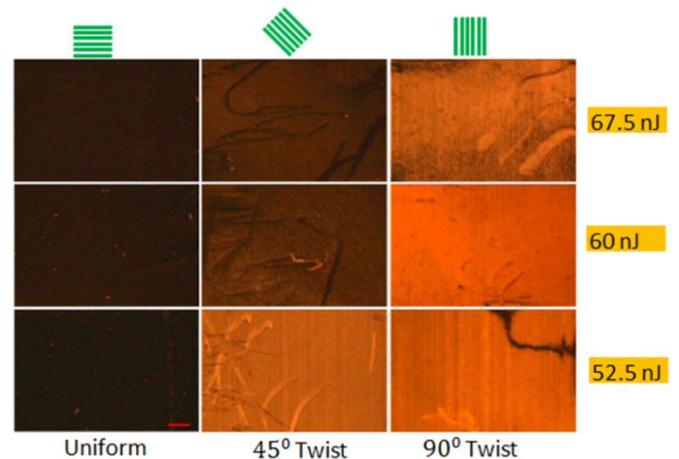

**Figure 7.** Transmission results of the nine different regions observed by polarized microscope from figure 5. Each column represents different LC alignment mode written at different pulse energies by fs laser along each row as illustrated in figure

2a. Red bar represents 0.1mm and green bars represents alignment direction of the nanogrid patterns.

As already mentioned, the columns represent different alignment directions (green bars), where each row was written at different fs laser intensities, which in turn leads to different retardation and threshold voltage levels, as its influence on phase retardation by means of an ITO thickness. The first column (from left) describes an anti-parallel\parallel (PA) LC cell. As can be seen, in the region that was written with 67.5 nJ (PA-67.5nJ), the molecules start orienting at relatively low threshold voltage (0.5±0.1V), which is smaller than the expected threshold voltage for LC E-7. The PA-60nJ region exhibits a higher threshold voltage (~0.9V), while the PA-52.5nJ region did not respond to the applied voltage for all alignment regions, due to ITO nanogrid discontinuity, that occurs during the fabrication process.

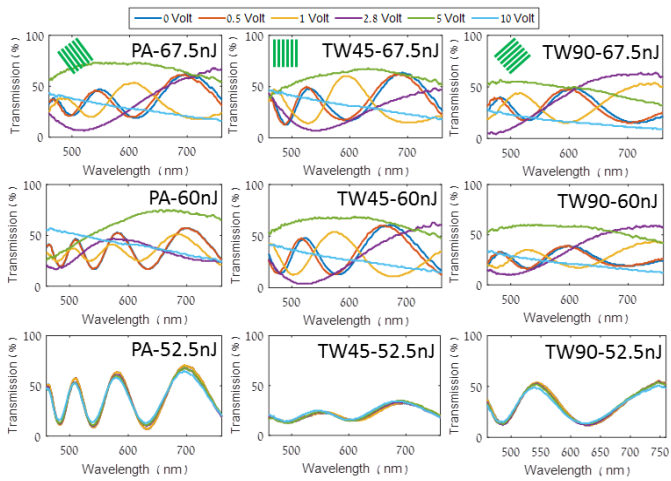

**Figure 8.** Spectrometer readout. Transmission spectra for various alignment regions at different applied voltages (sine wave at 1 kHz) between crossed polarizers. The green bars represent the orientation of nanograting patterns.

The apparent lower threshold value may indicate the existence of some pre-tilt angle and slight variation of it on the nanogrid surface because the nanogrid is not completely uniform. In addition, as expected at the high voltage limit (<~5 V), the transmission of the various alignment regions is much less dependent on the wavelength in the whole range of the visible spectrum. For the $45^0$ twisted nematic LC cells (center column), both regions (TW45-67.5nJ and TW45-60nJ), exhibit low threshold voltage, that is also not typical for twisted nematic LC cell with identical characteristics. At the last configuration (right column), there is $90^0$ twist, thus the incoming linear polarized light will be rotated by 90 degree by the time it exits the opposite substrate. As in the previous cases, the threshold voltage seems to be relatively low compared to typical $90^0$ twisted nematic cell. Again this can be due to some pre-tilt angle and its inhomogeneity on the nanogrid, which also cause some decrease of the contrast.

To describe more clearly the threshold voltage of different alignment regions, the applied voltage was scanned in steps of 0.01V, using a function generator controlled through a LabVIEW designed software. Figure 9 shows the transmission versus applied voltage relationship for different alignment regions. According to Freedericksz threshold transition [1], we can define the threshold voltage for the splay and twisted nematic cells (without the chiral dopant), respectively as:

$$V_c^{splay} = \pi \sqrt{\frac{K_{11}}{\varepsilon_0 \Delta\varepsilon}} \quad (1)$$

$$V_c^{twisted} = \pi \sqrt{\frac{K_{11}}{\varepsilon_0 \Delta\varepsilon}} \left[1 + \frac{(K_{33} - 2K_{22})}{K_{11}}\left(\frac{\phi}{\pi}\right)^2\right]^{\frac{1}{2}} \quad (2)$$

Where d is the LC thickness, $K_{11}$ is a splay elastic constant, $\Delta\varepsilon$ is a dielectric anisotropy, $\phi$ is the twist angle and $\varepsilon_0$ is the vacuum permittivity. For LC E7: $\Delta\varepsilon = 14$, $K_{11} = 10.8^{-12}N$, $K_{22} = 6.8^{-12}N$ and $K_{33} = 17.5^{-12}N$ at $20^0 C$ [20]. Thus the expected theoretical threshold voltage is: 0.92V for uniform PA cells, 0.96V and 0.93V for $90^0$ and $45^0$ twisted nematic cells, respectively. As can be seen from the interpolant curves in figure 9, only the anti-parallel\parallel region written at 60 nJ (PA-60nJ), close to the expectations, while the other regions have a relatively lower threshold voltage.

Low threshold voltage indicates on weaker anchoring energy, which is characterized, inter alia, by low relaxation time. For measuring the relaxation time, a collimated laser diode with a wavelength of 633nm was used as a light source, while the LC device was placed under crossed polarizers oriented at $45^0$ with respect to the polarizer axis.

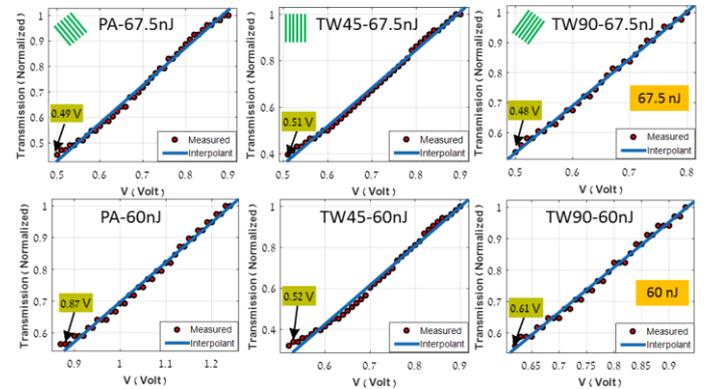

**figure 9.** Threshold voltage of different alignment regions deduced by interpolation curves measured by the optical setup illustrated in figure 6.

A sinusoidal 1 kHz with 1Vamp wave modulated by a low frequency square wave, generated from an arbitrary wave form generator, was applied to the device. The transmitted intensity was monitored with a photodetector connected to an oscilloscope. A summary of the 10%-90% fall time curve for a planar alignment region written at 60 nJ is shown in figure 10.

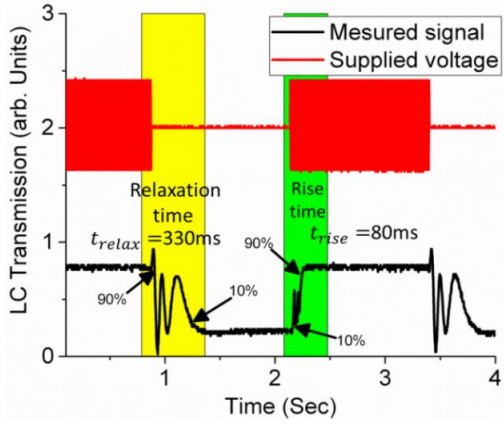

**Figure 10.** Measured time response of a planar alignment region written at 60 nJ measured between crossed polarizers, while the device oriented at 45-degree with respect to the polarizer axis. The transient oscillations are the result of relatively thick LC cell.

According to dynamics of the Freedericksz transition in splay geometry under the one splay elastic constant approximation, the relaxation time is given by:

$$t_{relax} = \frac{\gamma_\theta}{K_{11}}\left(\frac{d}{\pi}\right)^2, \quad t_{rise} = \frac{t_{relax}}{\left(V/V_c^{splay}\right)^2 - 1} \quad (3)$$

Where $\gamma_\theta$ is the rotational viscosity of the liquid crystal. For E-7 $\gamma_\theta = 243m\,[Pas \cdot S]$ [21], therefore the estimated relaxation time should be about 294.2ms for LC thickness of d=11.36µm and a rise time of 68.6ms at V=2V. Based on the experimental results the measured relaxation time is $t_{relax}$=330ms and the rise time is $t_{rise}$=80ms. Despite the small deviation between the measured and calculated values, the measured relaxation time is still in plausible range for thick LC devices as the Equations (3) are valid for small perturbation. Furthermore, there are few options that can be done to decrease the time response. First, smaller thickness of the liquid crystal layer will lead to better switching times. Second, substitution of the liquid crystal to ferroelectric, nano-polymer stabilized blue phases or LC materials with low viscosity will also decrease the response times. Third, we believe, that by further optimization of this process it will be possible to improve the switching times by increasing the pretilt angle during nanograting manufacturing. Finally, such driving techniques as voltage overshooting and undershooting [22] can be utilized in order to reduce the response times.

This fact makes us believe that the device is in the anti-parallel alignment geometry which is expected to be slower than the parallel alignment mode. This is expected because there is no reason for the molecules to prefer being in one direction of a nanogroove but not the opposite direction; however the constraint imposed by the top substrate alignment forces the molecules to align in antiparallel to minimize the free energy. In order to evaluate the anchoring energy ($W_A$) between the nematic LC molecules and the nanogrid substrate we used the improved high electric field method, proposed by Nastishin et al. [23]. According to this method, the anchoring energy can be calculated from a linear fit for the voltage (V-V') vs phase retardation normalized by the zero voltage phase Γ(V-V')/Γ₀ dependence. Where V'=α(Δε/ε∥)$V_C^{splay}$ and α coefficient varies between 1-2/π, and defined with all the details of the algorithm in [23].

The measured anchoring energy from PA-60nJ alignment region was on the order of $1.063 \times 10^{-6}\,J/m^2$, as demonstrated in figure 11. This anchoring energy is comparable to that measured using photo-aligning polymer materials, which is usually in the range of $10^{-5} - 10^{-6}\,Jm^{-2}$ [6-7].

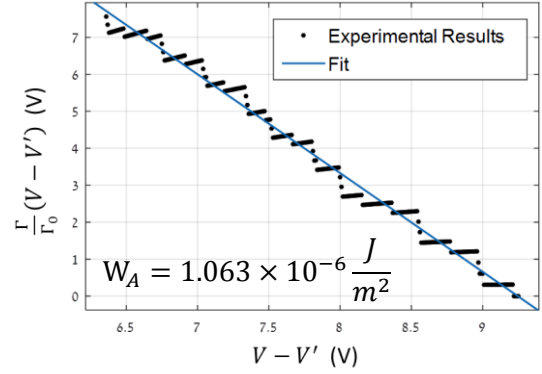

**Figure 11.** Plot of Γ(V-V')/Γ₀ vs V-V' and linear fit used for anchoring energy evaluation ($W_A$) for PA-60nJ alignment region, measured by electric field techniques[23], yielding the value of $1.063 \times 10^{-6}\,J/m^2$.

Another major quality describing parameter is the contrast ratio. To evaluate the contrast ratio, the device was placed between crossed polarizers at $45^0$ with respect to the polarizer axis. The transmission versus voltage was measured by the photodetector, while the He-Ne laser was used as light source. The results presented in figure 12, show contrast value of 30:1 for the TW45-60nJ alignment region. As was expected from figure 7, the contrast ratio is relatively low. On the one hand it could be referred to the nanoscale inhomogeneity of the nanograting surface; on the other hand, the PA-52.5nJ region from Figure 7, exhibits a good dark state, which indicates that the contrast ratio can be increased during further optimization process. The nanogratings patterns consist of alternating regions of 145nm ITO lines and glass substrate (or few nanometer thickness residual ITO).

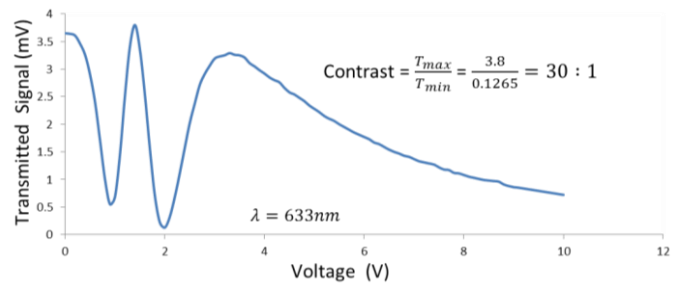

**Figure 12.** Voltage dependence of transmitted signal of TW45-60nJ liquid crystal region. The transmission was measured by photodetector with He-Ne laser light source, while the LC

device was placed between crossed polarizers oriented at $45^0$ with respect to polarizer axis. The deduced contrast ratio is 30:1.

Thus, the transmittance of incident light is expected to be higher than those of a uniform ITO layer in typical liquid crystal devices. As a proof of principle, we measured the transparency at the center grid region and at the outside area as depicted in figure 13. The results show around 3% higher transmittance at the TW45-60nJ grating region starting above the wavelength of 650nm and about 10% higher transmittance through the LC device with two ITO nanogrid substrates written under fs laser pulse energy of 0.76µJ and 0.68µJ respectively. The transmittance of the LC device with two uniform ITO substrates is around 84% at 650nm as illustrated in figure 13, meaning each interface is transmitting 92%. The LC device with one nanogrid side has transmittance of 87%, meaning the nanogrid interface is transmitting 95%; hence the LC device with two nanogrids is supposed to transmit around 90%.

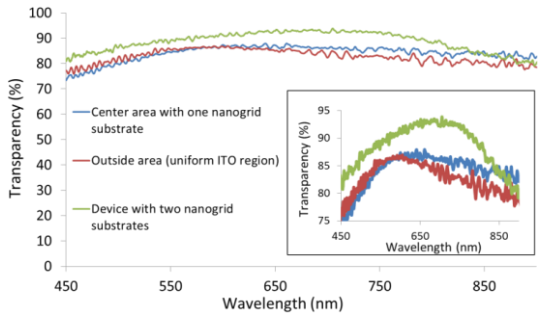

**Figure 13.** Transparency measurement without polarizers of the second LC device with two nanogrid substrates and central region of LC device with one nanogrid substrate (TW45-60nJ), compared to uniform ITO region.

In reality, the transmittance of LC device with two nanogrid substrates is 94%, meaning an improvement of 10% compared to LC device with two uniform ITO substrates, which can come partially from the fact that this large area nanogrid has smoother surface so it scatters less.

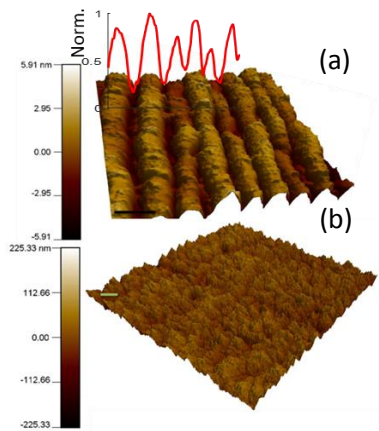

**Figure 14.** AFM Roughness measurement. (a) Nanogrid structure and normalized averaged roughness (red curve). (b) Uniform ITO layer. The black and green bars represent $0.25\ \mu m$ and $0.5\mu m$, respectively.

This is expected as this grating was written with weaker focusing which is known to give smoother modification, less cracks and the gaps between the lines are less visible (figure 14a). The slight decrease of the transparency at lower wavelengths maybe attributed to the relatively larger scattering due to nanoscale inhomogeneities of the grid surface. It is important to note once again that the ITO nanogrid demonstrated in figure 14a was created directly on ITO film (figure 14b) by fs laser.

For transmission analysis of nanogrid substrate we used 4x4 matrix method and form birefringence theory [24-25]. The periodic gratings should behave as negative index uniaxial film with optic axis along the gratings vector. In terms of refractive indices along the grating ($n_o^g$) and perpendicular to it ($n_e^g$), we can write equations 4-7 up to the second-order expressions [25]:

$$n_{TM_0} = \frac{n_m \cdot n_g}{\sqrt{n_g^2(1-f) + f \cdot n_m^2}} \quad (4)$$

$$n_{TE_0} = \sqrt{n_m^2(1-f) + f \cdot n_g^2} \quad (5)$$

$$n_o^g = n_{TE_2} = \left\{ n_{TE_0}^2 + \frac{1}{3}\left[\frac{\pi(1-f)p}{\lambda}\right]^2 \left(n_g^2 - n_m^2\right)^2 \right\}^{1/2} \quad (6)$$

$$n_e^g = n_{TM_2} = \left\{ n_{TM_0}^2 + \frac{1}{3}\left[\frac{\pi(1-f)p}{\lambda}\right]^2 \left(\frac{1}{n_g^2} - \frac{1}{n_m^2}\right)^2 n_{TM_0}^6 n_{TE_0}^6 \right\}^{1/2} \quad (7)$$

Here $n_g$ is the refractive indexes of the grid, $n_m$ is the refractive indexes of the substrate materials, $n_{TEo}$ and $n_{TMo}$ are the zero order terms of refractive index for light polarized along the grating and perpendicular to it, respectively and f is a fill factor defined as the ratio between the width of the grid (w) and the period (p) as illustrated in figure 15.

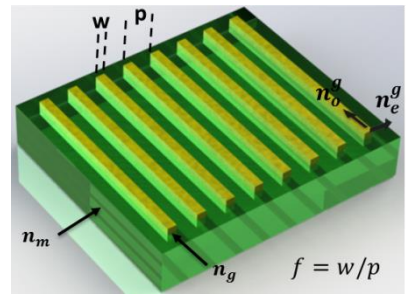

**Figure 15.** The sketch of the femtosecond laser induced nanogratings. W represents the width of the nano-bar and p is the pitch of the nanograting. The refraction index of the grating patterns and the substrate are labeled as $n_g$ and $n_m$,

respectively. $n_o^g$ represents refractive index along the grating and $n_e^g$ – perpendicular to it.

From a simple calculation of the optics of subwavelength structures, it is known that they can act as antireflection coatings with careful choice of the structure parameters. Equations 4-7 used for modeling the uniaxial plate within the 4x4 matrix approach [24]. Figure 16 shows the transmission as a function of wavelength for fill factor f=0.8 and ITO layer thickness of 145 nm with ITO dispersion relation based on fit and given by: $n_{Dis}$=1.7921+0.40766/$\lambda$-0.45655/$\lambda^2$ +0.27672/$\lambda^3$ -0.074223/$\lambda^4$+0.0080674/$\lambda^5$+i(0.016005-0.031929/$\lambda$- -0.026768/$\lambda^2$-0.010387/$\lambda^3$+0.0019364/$\lambda^4$-0.00013227/$\lambda^5$). Although the dispersion relations for ITO are sensitive to the preparation process and can vary from one manufacturer to another, the purpose of Figure 16 is to show that the transmission of the patterned structure is higher than the continuous film for both TE (Tss) and TM (Tpp) polarizations. We have checked that this is the case for other available dispersion relations of ITO. Note that for the continuous TE and TM curves coincide at normal incidence.

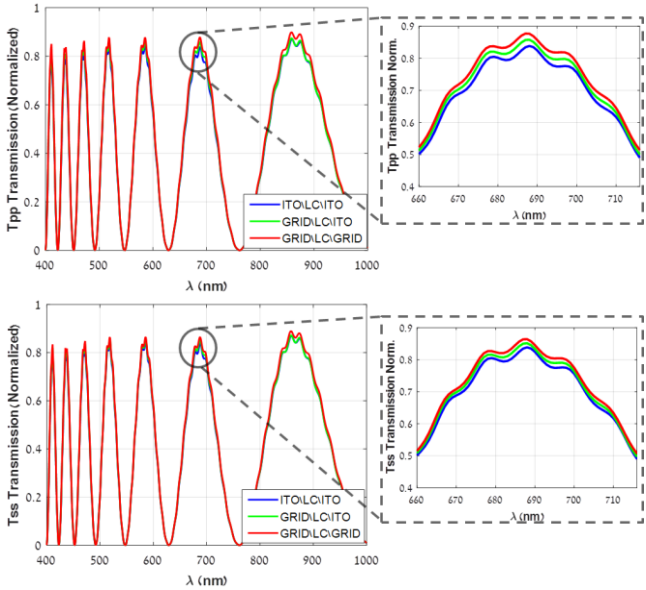

**Figure 16.** Calculated transmission of Tpp and Tss, for three different substrates configuration of LC device with d=11.36µm, by the 4x4 matrix method. LC between two uniform ITO substrates (ITO\LC\ITO), LC between uniform ITO and nanogrid substrates (Grid\LC\ITO) and LC between two nanogrid substrates (Grid\LC\Grid). The transmission for Grid\LC\Grid configuration is higher by almost 5% through all the spectra as demonstrated in the range 660nm-720nm.

In order to validate the transmission spectrum at different alignment regions, we used Jones Matrix method. The expression for outgoing light in the case of PA (while we assume a small twist angle created during the construction process), can be described by Jones calculus as:

$$V_{out} = P_A \cdot R(-\zeta) \cdot W_{Grid} \cdot R(\zeta) \cdot R(-\psi) \cdot R(-\phi_{LC}) \cdot W_{Twist} \cdot R(\psi) \cdot P_P \quad (8)$$

Where $W_{Twist}$ is Jones matrix for the optical transmission of a general twisted nematic cell and $W_{Grid}$ represents the matrix of the nanograting. Both matrices can be written as:

$$W_{Grid} = e^{-i\vartheta} \begin{pmatrix} e^{-i\Gamma_{Grid}/2} & 0 \\ 0 & e^{i\Gamma_{Grid}/2} \end{pmatrix} \quad (9)$$

$$W_{Twist} = R(-\phi_{LC}) \begin{pmatrix} \cos x - i\frac{\Gamma_{LC}}{2}\frac{\sin x}{x} & \phi_{LC}\frac{\sin x}{x} \\ -\phi_{LC}\frac{\sin x}{x} & \cos x + i\frac{\Gamma_{LC}}{2}\frac{\sin x}{x} \end{pmatrix} \quad (10)$$

Where the $\varphi_{LC}$ is the LC twist angle and $x$ is given by:

$$x = \sqrt{\phi_{LC}^2 + \left(\frac{\Gamma_{LC}}{2}\right)^2} \quad (11)$$

The phase factor $e^{-i\vartheta}$ can be neglected and P$_A$, P$_P$ represent analyzer and polarizer matrix and vector, respectively, given by:

$$P_A = \begin{pmatrix} 0 & 0 \\ 0 & 1 \end{pmatrix}, \quad P_p = \begin{pmatrix} 1 \\ 0 \end{pmatrix} \quad (12)$$

R(ζ) and R(ψ) perform coordinate system rotation of the nanogrid and the LC molecules and both defined as:

$$R(\psi) = \begin{pmatrix} \cos\psi & \sin\psi \\ -\sin\psi & \cos\psi \end{pmatrix}, R(-\psi) = \begin{pmatrix} \cos\psi & -\sin\psi \\ \sin\psi & \cos\psi \end{pmatrix} \quad (13)$$

The LC phase retardation ($\Gamma_{LC}$) can be written as:

$$\Gamma_{LC} = \frac{2\pi}{\lambda}\int_0^d \left(n_{eff}(z) - n_o\right)dz \quad (14)$$

Where λ is the wavelength, $n_0$ and $n_{eff}(z)$ are the ordinary and the extraordinary refractive index of the LC, respectively and defined as:

$$n_{eff}(z) = \frac{n_\perp n_\parallel}{\sqrt{n_\perp^2 \cos^2\theta(z) + n_\parallel^2 \sin^2\theta(z)}}, n_o = n_\perp \quad (15)$$

For LC E-7 $n_\parallel$=1.693 and n$_\perp$=1.499. The extraordinary refractive index depends on the angle θ(z) and can be evaluated by second dynamic equation as follows [26]:

$$\frac{(1+k\sin^2\theta(z))}{\pi^2}\frac{d^2\theta(z)}{dz_r^2} + \frac{k\sin\theta(z)\cos\theta(z)}{\pi^2}\left(\frac{d\theta(z)}{dz_r}\right)^2 + V_r^2 \sin\theta(z)\cos\theta(z) = 0, \quad (16)$$

Where $k = (k_{33} - k_{11})/k_{11}$, and $V_r$ is the voltage normalized to the threshold. In the case of twisted nematic alignment regions the outgoing light can be described by equation 8 with appropriate twist parameters.

The measurement and the simulation results for PA and twisted nematic of $45^0$ and $90^0$ at zero bias are demonstrated at Figure 17. Since the nano patterns are not ideal, we averaged and optimized the calculations over several domains, having slightly different pretilt angle and fill factor, in order to achieve the good fit between the experimental and calculated curves.

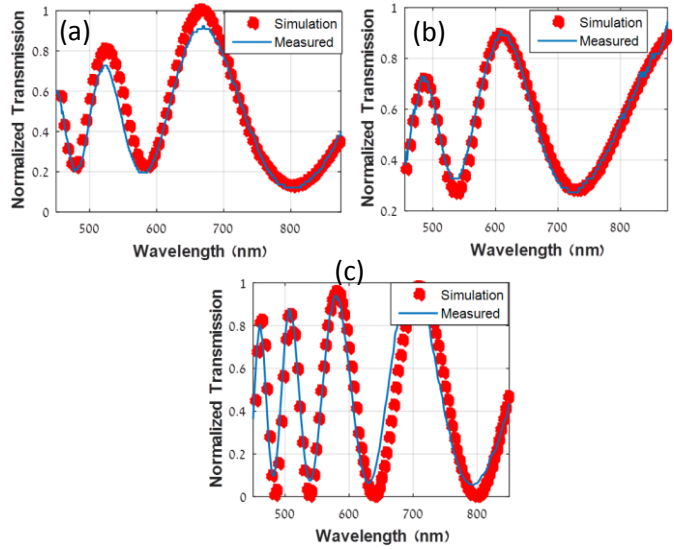

**Figure 17.** Simulated and experimental data measured at zero bias for $45^0$ twist at TW45-60nJ (a), $90^0$ twist at TW90-67.5nJ (b), and parallel alignment with $0.2^0$ twist at PA-52.5nJ (c).

The second large area device with two ITO nanogrid substrates was examined at various supplied voltages with the same microscope as used in the experiment of figure 5, but in the reflection mode as illustrated in figure 18.

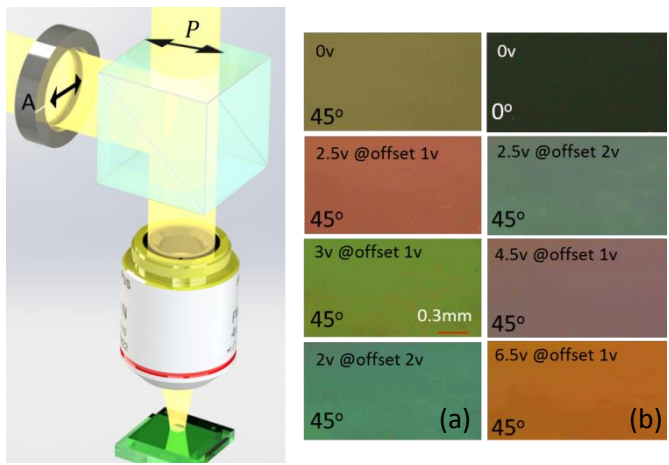

**Figure 18.** Illustration of reflection from ITO nanogrid LC device under polarized microscope with crossed polarizers. (a) Linear polarized light (P) reflected from the device, oriented at 45-degree with respect to polarizer axis. The reflected light reaches the analyzer (A) through the beam splitter (only for illustration purpose) toward the camera. (b) The device was examined under various applied voltages. The red bar represents 0.3mm.

The reflectivity results reveal a uniform extinction, when the LC device optical axis is parallel to the axis of one of the polarizers ($0^o$). Furthermore, when the device is placed with its optical axis $45^0$ from the polarizer's axis, it exhibits a uniform reflection change as function of voltage.

In addition, the transmission spectrum through the device, placed at $45^0$ between crossed polarizers, was measured for normal incidence at different voltages as illustrated in figure 19.

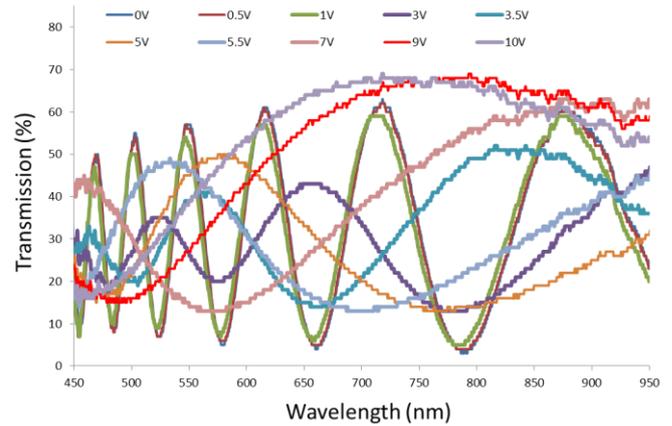

**Figure 19.** Spectrometer readout. Transmission spectra of second LC device with large nanogrid area at different applied voltages (sine wave at 50Hz).

Upon the threshold voltage of 1V, the transmission spectrum considerably changed, as shown in figure 19. Figure 18 and figure 19 definitely indicates that two ITO nanogrid substrates can be used for LC alignment purposes with larger area and higher transmission.

To examine the stability of the device, we checked the contrast variations during 8 hours of continuously biased device under polarized microscope for two different orientation modes. At the first mode, the device alignment direction was rotated at 45° relative to the optical axis of the polarizer (figure 20a). At the second mode the device was placed at the extinction position with its alignment direction parallel to the polarizer axis (figure 20b). In both cases the device was biased under 7V, 80Hz and offset voltage of 1V, while the images were captured every hour from the moment we applied the voltage. Then, the contrast was calculated by: $(I_{max} - I_{min})/(I_{max} + I_{min})$, where $I_{max}$ and $I_{min}$ represent the averaged values of images in Figure 20a and figure 20b, respectively, taken at appropriate times. Finally, we calculated the variation of the contrast at each hour, compared to the contrast received at the the moment of turning it on. As can be seen from figure 20c, the contrast variations are negligible (<1%) throughout the test period and the small variations could simply be part of the system instability during 8 hours.

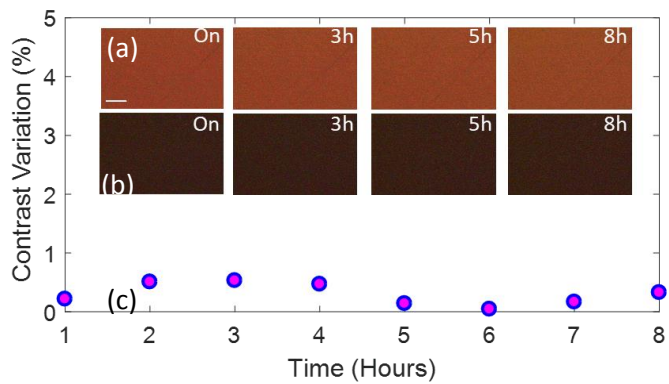

**Figure 20.** Stability test of the the device prepared using two substrates with ITO nanogrids under polarized microscope with crossed polarizers. The images were captured at 45° (a), and the extinction (b), positions relative to polarizer axis each hour during the test. (c) Analysed contrast variation from the moment we turned on the voltage. The bar represents $125\mu$.

The pretilt angle $\alpha_P$ of the second device was determined by the classical crystal rotation method [27]. The device has been rotated by a stepping motor, which was controlled through the LabVIEW program, while the transmitted laser diode light ($\lambda = 635 nm$) was collected by the power meter. Figure 21 shows the experimental results and the fitted curve, which was used to determine the pretilt angle of $\alpha_P = 0.48°$.

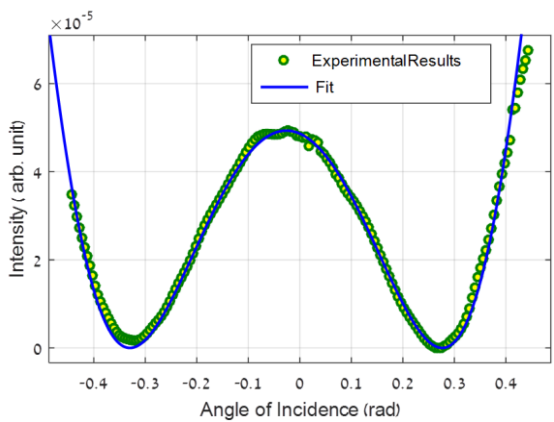

**Figure 21.** Measured Intensity as a function of incidence angle for the second nanogrid LC device (with two nanopatterned ITO substrates). Pretilt angle of $\alpha_P = 0.48°$ was deduced by fitting between the measured and the calculated transmission intensity.

## 4. Conclusions

Nanogrid patterns created directly on ITO coated substrates by femtosecond laser, can be used as alternative method for LC alignment with higher transparency, without any additional alignment layers. The LC devices exhibit low threshold voltage, but at the same time reasonable response time of 330ms, suitable for 11.36 μm thick anti-parallel alignment LC devices. The anchoring energy that was measured for the PA-60nJ alignment region was $1.063 \times 10^{-6}\ J/m^2$, comparable to the anchoring energy of nematic LC on photosensitive polymers. However, the contrast ratio at this stage is moderate, we believe it can be increased by optimization of the intensity of the fs laser during the fabrication process of the nanogrid patterns as well as the assembly process of the LC device. In addition, we show that this periodic ITO structure increases the transmission through the LC device by 3% in the case, where it is assembled with a second uniform ITO substrate and about 10% with two nanogrid ITO substrates written under fs laser pulse energy of 0.76μJ and 0.68μJ respectively. These results pave the way for additional research and optimizations of the process in order to make this method good alternative for typical industrial LC alignment techniques and for creating higher transparency electrodes. As LC displays energy costs are higher by around 30% than light emitting diode displays, particularly for large areas, the results presented here have the potential to revamp large area LC displays as more cost effective solutions.

## Conflicts of interest

There are no conflicts to declare.

## Acknowledgements

This research is supported partially by the National Research Foundation, Prime Minister's Office, Singapore under its Campus for Research Excellence and Technological Enterprise (CREATE) program, and Engineering and Physical Sciences Research Council (EPSRC) (EP/M029042/1). The data for this work is accessible through the University of Southampton Institutional Research Repository.
doi: 10.5258/SOTON/D0199.